\documentclass{article}
\usepackage{tabularx}
\usepackage[acronym,toc]{glossaries}

\usepackage[english]{babel}
\usepackage{amsmath}
\usepackage{amsfonts}
\usepackage{amssymb}
\usepackage{amsmath}
\usepackage{amsfonts}
\usepackage{amssymb}
\usepackage{amsfonts}
\usepackage{mathtools}

\usepackage{listings}
\usepackage{tikz-uml}

\usepackage{amsthm}

\usepackage{fmtcount}
\usepackage{graphicx}
\usepackage{caption}
\usepackage{subcaption}

\usepackage{courier}
\usepackage[ruled,vlined]{algorithm2e}

\usepackage[usenames,dvipsnames]{pstricks}
\usepackage{pst-grad} 
\usepackage{pst-plot} 

\usepackage[acronym,toc]{glossaries}
\usepackage{authblk}

\graphicspath{{img/}}

\usepackage{color}
\definecolor{Red}{rgb}{1,0,0}
\definecolor{mygreen}{rgb}{0,0.6,0}
\definecolor{mygray}{rgb}{0.3,0.3,0.3}
\definecolor{mylightgray}{rgb}{0.95,0.95,0.95}
\definecolor{mymauve}{rgb}{0.58,0,0.82}
\definecolor{mywhite}{rgb}{1,1,1}
\newacronym{FS}{FS}{File System}
\newacronym{SG}{SG}{Smart Grid}
\newacronym{UC}{UC}{Ubiquitous Computing}
\newacronym{OS}{OS}{Operating System}
\newacronym{AO}{AO}{Artificial Organism}

\newacronym{UML}{UML}{Unified Modelling Language}
\newacronym{VM}{VM}{Virtual Machine}
\newacronym{OPC}{OPC}{Organic Processing Cells}
\newacronym{MOSFET}{MOSFET}{Metal Oxide Semiconductor Field Effect Transistor}
\newacronym{CMOS}{CMOS}{Complementary Metal Oxide Semiconductor}

\newacronym{SuOC}{SuOC}{System under Observation and Control}
\newacronym{SOC}{SOC}{System on Chip}

\newacronym{CPU}{CPU}{Central Processing Unit}
\newacronym{API}{API}{Application Programming Interface}
\newacronym{IPV6}{IPv6}{Internet Protocol version 6}

\newacronym{SMAP}{sMAP}{Simple Measurement and Actuation Profile}
\newacronym{DPWS}{DPWS}{Devices Profile for Web Services}
\newacronym{RDF}{RDF}{Resource Description Framework}
\newacronym{RESTFUL}{RESTFUL}{REpresentational State Transfer}

\newacronym{FIPA}{FIPA}{Foundation for Intelligent Physical Agents}
\newacronym{ACL2}{ACL}{Agent Communications Language}
\newacronym{OSGI}{OSGi}{Open Services Gateway initiative}
\newacronym{HCI}{HCI}{Human Computer Interface}

\newacronym{WSAN}{WSAN}{Wireless Sensor and Actor Networks}
\newacronym{OC}{OC}{Organic Computing}
\newacronym{AMI}{AmI}{Ambient Intelligence}
\newacronym{ECA}{ECA}{Event-Condition-Action}

\newacronym{IOT}{IoT}{Internet of Things}
\newacronym{HIS}{HIS}{Home Information System}

\newacronym{DOS}{DoS}{Denial of Service}
\newacronym{AES}{AES}{Advanced Encryption Standard}

\newacronym{SOA}{SOA}{Service Oriented Architecture}
\newacronym{WOA}{WOA}{Web Oriented Architecture}
\newacronym{ROA}{ROA}{Resource Oriented Architecture}

\newacronym{REST}{REST}{REpresentational State Transfer}
\newacronym{HTTP}{HTTP}{Hyper Text Transfer Protocol}
\newacronym{URI}{URI}{Unified Resource Identifier}
\newacronym{SSL}{SSL}{Secure Socket Layer}
\newacronym{COAP}{CoAP}{Constrained Application Protocol}
\newacronym{SOAP}{SOAP}{Simple Object Access Protocol}
\newacronym{JSON}{JSON}{JavaScript Object Notation}
\newacronym{XML}{XML}{Extensible Markup Language}

\newacronym{IP}{IP}{Internet Protocol}
\newacronym{ICMP}{ICMP}{Internet Control Message Protocol}
\newacronym{QOS}{QoS}{Quality of Service}
\newacronym{VPN}{VPN}{Virtual Personal Network}
\newacronym{DNS}{DNS}{Dynamic Name Server}
\newacronym{SNMP}{SNMP}{Simple Network Management Protocol}
\newacronym{UPNP}{UPnP}{Universal Plug and Play}
\newacronym{WSN}{WSAN}{Wireless Sensor and Actor Networks}
\newacronym{LAN}{LAN}{Local Area Network}

\newacronym{6LOWPAN}{6LoWPAN}{IPv6 LoW Power Wireless Area Networks}
\newacronym{BER}{BER}{Border Edge Router}
\newacronym{RNDIS}{RNDIS}{Remote Network Driver Interface Specification}
\newacronym{ACL}{ACL}{Access Control List}

\newacronym{EMS}{EMS}{Energy Management System}
\newacronym{RFID}{RFID}{Radio Frequency IDentification}
\newacronym{FPGA}{FPGA}{Field-Programmable Gate Array}
\newacronym{BLE}{BLE}{Bluetooth Low Energy }

\newacronym{EMMA}{EMMA}{Environment Monitoring and Management Agent}
\newacronym{NPN}{NPN}{Numerical Petri Network}
\newacronym{MCKP}{MCKP}{Multiple Choice Knapsack Problem}
\newacronym{MKP}{MKP}{Multiple Knapsack Problem}
\newacronym{PBO}{PBO}{Pseudo Boolean Optimization}

\newacronym{ANC}{ANC}{Artificial Neural Controller}
\newacronym{ANN}{ANN}{Artificial Neural Network}
\newacronym{NWSN}{NWSN}{Neural Wireless Sensor Network}
\newacronym{NAC}{NAC}{Network Average Consensus}
\newacronym{MLP}{MLP}{Multi-Layer Perceptron}

\newacronym{OWA}{OWA}{Ordered Weighted Average}
\newacronym{VP}{VP}{Voting Procedures}
\newacronym{MAS}{MAS}{Multi Agent System}
\newacronym{CS}{CS}{Consensus Seeking}
\newacronym{DS}{DS}{Dynamic System}
\newacronym{CT}{CT}{Control Theory}
\newacronym{DT}{DT}{Decisions Theory}
\newacronym{MNAC}{MNAC}{Multi Network Average Consensus}

\newacronym{ARM}{ARM}{Active Resource Middleware}
\newacronym{DA}{SC}{Service Choreography}
\newacronym{NS}{NS}{Name Space}
\newacronym{DNA}{DNA}{Dynamic Network Agent}
\newacronym{RNA}{RNA}{Residual Network Agent}

\newacronym{ES}{ES}{Embedded Systems}
\newacronym{AI}{AI}{Artificial Intelligence}
\newacronym{RE}{RE}{Responsive Environments}

\newacronym{MAC}{MAC}{Medium Access Control}
\newacronym{TDMA}{TDMA}{Time Division Multiple Access}
\newacronym{RPL}{RPL}{Routing Protocol for Low power and Lossy Networks}
\newacronym{DAG}{DAG}{Directed Acyclic Graph}
\newacronym{TUNSLIP}{TunSLIP}{Tunnel Serial Line Internet Protocol}
\newacronym{WLAN}{WLAN}{Wireless Local Area Network}

\newacronym{IETF}{IETF}{Internet Engineering Task Force}
\newacronym{IEEE}{IEEE}{Institute of Electrical and Electronics Engineers}

\newtheorem{definition}{Definition}

\begin{document}
\title{EMMA: A Resource Oriented Framework\\ for Service Choreography over\\  Wireless Sensor and Actor Networks}
\author[1]{Duhart Clement}
\author[1]{Sauvage Pierre}
\author[2]{Bertelle Cyrille}
\affil[1]{Department of Computer Science, LACSC, ECE Paris, France, \url{<duhart,sauvage>@ece.fr}}
\affil[2]{LITIS, FR CNRS 3638, Le Havre University, France, \url{cyrille.bertelle@univ-lehavre.fr}}

\date{}
\maketitle

\begin{abstract}
Current \gls{IOT} development requires service distribution over \gls{WSN} to deal with the drastic increasing of network management complexity. Because of the specific constraints of \gls{WSN}, centralized approaches are strongly limited. Multi-hop communication used by \gls{WSN} introduces transmission latency, packet errors, router congestion and security issues. As it uses local services, a decentralized service model avoid long path communications between nodes and applications. But the main issue is then to have such local services installed on the desired nodes. \gls{EMMA} system proposes a set of software to deploy and to execute such services over \gls{WSN} through middleware based on Contiki OS. This contribution presents \gls{EMMA} middleware, methodology and tools used to determine efficient service mapping and its deployment.
\end{abstract}

\begin{keywords}
Internet of Things (IoT), Wireless Sensor Network (WSN), Service Choreography (SC),  Middleware, Mobile Agent and Petri Network
\end{keywords}

\glsresetall
\section{Introduction}

The development of the \gls*{IOT} has some emphasis about current scientist issues and future industrial applications. Among the \gls*{IOT} applications, the \gls*{RE} are a part of a novel technological area. \gls*{RE} aims to transform our daily environments into intelligent spaces. 
Historically the domotic systems were automata systems for controlling appliances. Nowadays the term of \gls*{RE} is referring to an environment connected to Internet. 
On one hand, the data collection is used by remote service providers to manage macro issues, i.e. the energy providers which have to regulate energy production.
And on the other hand, the different appliances are managed by different service companies over Internet. An alarm system can be monitored by distant security guardians and an health care system helps older people in their daily tasks.
A major challenge for the \gls*{IOT} is the sharing of a common network infrastructure between all current and future services. \gls*{WSAN} are used to connect locally the different appliances into a common  wireless mesh networks. Appliances get an Internet access through others appliances. \gls*{WSAN} has important constraints in terms of bandwidth, throughput and payload. The network protocols proposed for \gls*{WSAN}, such as ZigBee, were incompatible with \gls{IP}, the Internet standard.

Since the 90's, a lot of research works have experimented approaches around \gls{IP} to facilitate \gls{WSAN} incorporation into Internet. \gls{6LOWPAN} protocol has been standardized to use IPV6 over IEEE 802.15.4 (ZigBee)  developed for energy constrained devices  \cite{Dunkels2003, VasseurJean-PhilippeandDunkels2010, Ko}. An HTTP based application layer called \gls{COAP} is currently investigated to provide WEB based communication transactions \cite{Kovatsch2011}. This protocol provides mechanisms for webservice interfaces like \gls*{REST} or \gls{SOAP} \cite{Moritz2011}.
A lot of open discussions remain regarding software architectures for \gls*{RE}. Assuming that all appliances use a common network protocol, their applications stay heterogeneous which is addressed commonly  by a central system. However in such situation, packet congestion appears around the routers according to network depth. Hence a new approach called \gls*{DA} is emerging to distribute locally the data exchanges. But the data heterogeneity is not still managed in framework found in literature.

This paper focuses on methodology of design, analysis and deployment of such distributed services over \gls{WSN}.
Section \ref{sec:relatedwork} presents a literature review regarding the different programming approaches of middleware to focus on current limitations of future required \gls*{ROA}. The proposed framework \gls{EMMA} is introduced in Section \ref{sec:emma}. This framework facilitates service choreography by providing a flexible distributed Publish-Subscribe mechanisms over \gls*{COAP}. Efficient methodology to distribute such data exchanges is presented in Section \ref{sec:mapping-problem} according to node hosting capacity with the consideration of deployment processes. Section \ref{sec:experimentation} presents results about association mapping of this rules and discuss about efficient deployment process. Finally, Section \ref{sec:conclusion} concludes about \gls{EMMA} approach and future works.

\section{Related work}
\label{sec:relatedwork}
 A middleware is a piece of abstraction software which provides advanced functionalities of hardware and network engines to applications. It is composed at least of a network stack, a multi-task manager and the drivers. Hadim et al. \cite{Hadim2006} and Rahman et al. \cite{Rahman2006} detail the different challenges regarding the network (scalability, mobility and dynamic topology), design (hardware abstraction, resource awareness and modular programming) and data (aggregation, heterogeneity and quality of service). 
The design of middleware stays an intensive research domain because  its  technologies issues are inherent of large applications area. 
Rubio et al. \cite{rubio2007programming} proposes a classification of programming middleware for \gls*{WSAN}: the \textit{macro-programming} and the \textit{node-centric}.

The \textit{macro-programming} consider a \gls*{WSAN} like an integrated system. Each node is a processing unit which executes particular operations assigned at a macro-level. The literature provides three main different approaches of \textit{macro-programming} in which the system is considered indifferently like a cluster of processing units, a distributed database or a \gls*{MAS}. 
Kushwaha et al. \cite{Kushwaha2007} presents the OASIS architecture to design  applications composed of different tasks distributed over the \gls*{WSAN}. They exchange directly their data in order to build computation flows over the \gls*{WSAN}. These tasks and their interconnections are designed offline and deployed remotely from supervisors. Hence the operations are executed directly inside the \gls*{WSAN} but they are managed remotely.
Costa et al. \cite{Costa2006} proposes an architecture which considers the \gls{WSAN} like a distributed database. Instead of collecting all data into a database, the requests are directly transmitted by broadcasting over the \gls*{WSAN}. The nodes resolve locally the request in order to provide an aggregated response to the requester.  
Fok et al. \cite{Fok2006} and Hackmann et al. \cite{Hackmann2006} have developed another macro-programming approach based on mobile agents. Each node has a \gls*{VM} in order to execute soft-coded applications. An application is composed of different role based agents which exchange data and perform operations onto a virtual tupple space. In \gls*{MAS}, each agent tries to satisfy its goals under its constraints in order to reach a global stationary point. Hence new constraints or goals can be added on runtime by the addition of agents in the virtual tupple space. This strong uncoupling between the application and hardware levels facilitates dynamic and online reprogramming of the \gls*{WSAN} such as there is no global problem formulation. Moreover, Liu et al. \cite{Liu2011} show that this approach is useful to load-balance energy consumption over the \gls*{WSAN} using mobile agent moving over the nodes according to the residual energy repartition. 

The \textit{node-centric} design considers the nodes like autonomous devices connected to \gls*{WSAN}. In such approach, the application term is referring to the software embedded into the nodes. They  collaborate with others nodes or Internet services like a traditional distributed system. Hence the middleware provides mechanisms for networked applications including service discovery, protocol interfaces and data heterogeneity management. Dunkels et al \cite{dunkels2004contiki} has developed a complete micro-operating system able to execute such applications in parallel on a single node. These applications communicate with other local applications through an event-based messaging engine and remotely with others services with the uIP stack. They propose an advanced solution using standard protocol \gls*{6LOWPAN} to deploy remotely the binary applications over the air like on traditional computer systems. However in \gls*{WSAN}, the nodes can not be configured manually and individually according to the large scale of the network. Delicato et al \cite{Delicato2003}, Souto et al \cite{Souto2004} and Khedo et al. \cite{Khedo2009} propose mechanisms based on Publish-Subscribe pattern to configure remotely the data exchanges between the applications. The applications subscribe to a class of data through their middleware in order to receive them when they are published on the network. 

Both approaches have different interests according to the \gls*{WSAN} purposes. For example in data collection, a distributed database is more interesting than a \textit{node-centric} approach because the system is homogeneous and do not require to collect systematically all data on a central database. In case of \gls*{IOT} applications, the devices and the services are produced and added by different manufacturers along the network lifetime. The lack of standard middleware forces the manufacturers to use \textit{node-centric} middleware in their products. In such situation, the inherent heterogeneity of the  applications and the network communications should be managed by supervisors in order to maintain network consistency and to preserve its resources.  Kuorilehto et al. \cite{kuorilehto1900survey} conclude their survey that \textit{Currently, they implement technologies and algorithms for application distribution but lack an approach combining a distributing middleware layer to OS providing a single node control.} 
Recently, Cherrier and al. \cite{cherrier2014bec, cherrier2012services} has proposed a new framework based on choreography of services for \gls*{WSAN}. They combine both approaches proposing a distributed  \textit{node-centric} based middleware with a high level language to describe macro-applications. The nodes provide web-services to produce or consume data which can be used simultaneously by different applications. Hence the authors propose a model of Finite State Machine (FSM) in order to guarantee the system consistency according to the different network exchanges between the node applications. The authors compare their work with centralized approaches based on the orchestration of services in which a gateway collects all the data and controls remotely the system. They demonstrate analytically and empirically that choreography model has better performances in terms of reliability, energy efficiency and scalability than orchestration.
 However the resource limitations in term of memory on node do not allows them to manage the heterogeneity of sequential protocols used at application layer such as \gls*{SOAP}. Hence, Guinard et al \cite{guinard2010resource} proposes a \gls*{ROA} framework for service choreography. This \gls*{RESTFUL} model uses the Publish-Subscribe mechanisms at resource level.  Hence, the protocols which require sequential exchanges are implemented like a FSM of several Publish-Subscribe instead of a part of the webservice. An application is resumed by a graph of resource interactions described like Publish-Subscribe configurations. When a resource is changing, its contain is transmitted to its dependent resources in order to form a cascading computation flow over the \gls*{WSAN}. In literature review, the establishment of choreography at resource level between node applications are operated manually from supervisor. 
 
 \glsreset{ROA}
 In this paper, the following questions are addressed to propose a \gls*{ROA} middleware with self-reconfiguration mechanisms for service choreography over \gls*{WSAN}: 
 \begin{center}
\textit{ How to map automatically the resources over the \gls*{WSAN} ?\\ How to guarantee the consistency of their interactions ? \\And how to deploy them in a large scale network ?}
 
 \end{center}

\newpage
\section{EMMA Framework}
\label{sec:emma}

\gls*{EMMA} is a middleware for service choreography over \gls*{WSAN}. Its \glsreset{ROA}\gls*{ROA} encapsulates node services into containers in order to form a resource tupple space. Among the different services, a \gls*{VM} executes reactive agents which models an augmented Publish-Subscribe mechanism distributed over the \gls*{WSAN}. These agents have the ability to transcode \gls*{COAP} requests in order to manage locally the heterogeneity with other middlewares or remote services using \gls*{COAP}. Such as these agents are themselves resources, they have the capacity to be self-deployed with self-rewriting abilities.

A \gls*{DA} is a service choreography formed by a set of services connected by the agents. The resource tupple space provided by the middleware forms a complete abstraction between the service choreography and its execution supports on node. A remote supervisor is responsible to define the best mapping of resources to deploy \gls*{DA} in order to preserve nodes and network load. This mapping takes in consideration the deployment process which is itself a \gls*{DA} composed of agents.

This Section presents the middleware architecture with the different basic \gls*{EMMA} services of agent executions, system interfaces and data storages. The proposed graphical framework based on an augmented Petri Network provides an easy solution to design complex \gls*{DA} with composition features. Its use allows mathematical background to be reused in order to analysis event diffusion of the \gls*{DA} in order to guarantee its consistency. Finally, the different strategies of resource deployment are presented with peer-to-peer deployment, composed deployment and self-deployment.

\subsection{Middleware}

\subsubsection{Resource Abstraction}
\begin{figure}[h!]
	\begin{subfigure}[b]{0.4\textwidth}
		\centering
		\includegraphics[height=6cm]{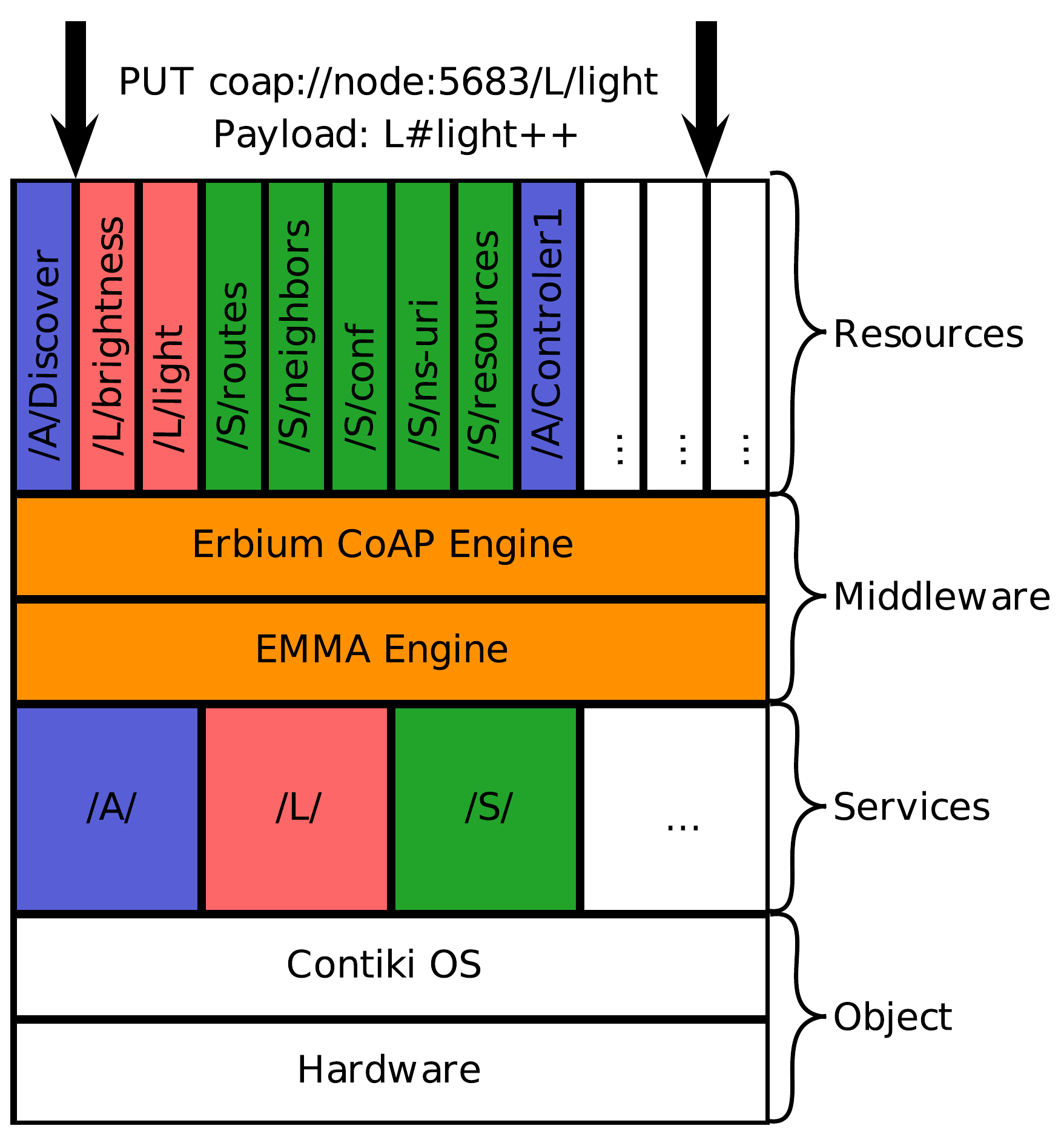}
		\caption{\label{fig:middleware}Middleware architecture}
	\end{subfigure}
	\hspace{0.8cm}
	\begin{subfigure}[b]{0.4\textwidth}
		\centering
		\includegraphics[height=6cm]{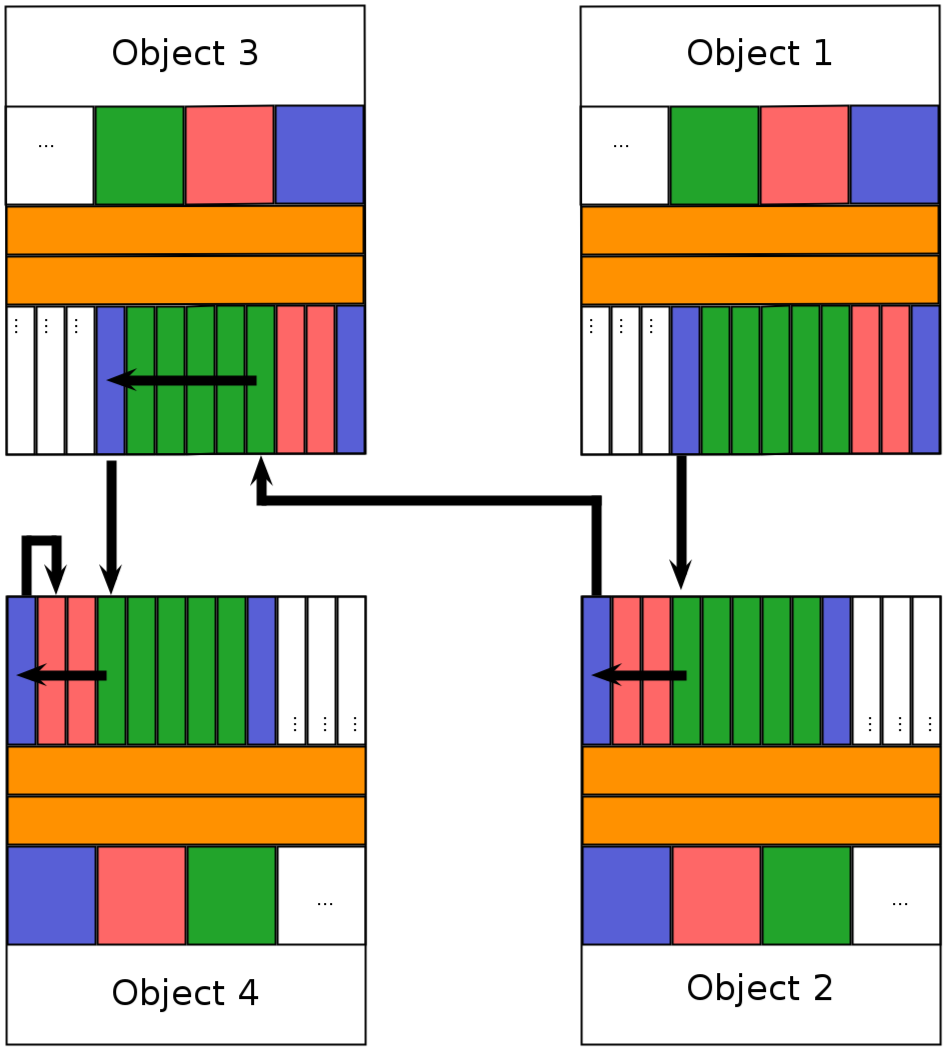}
		\caption{\label{fig:ex-choreography}Service choreography example}
	\end{subfigure}
	\caption{The middleware provides an abstraction between hard-coded service of the node and their reprogrammable choreography by EMMA agents.}
\end{figure}

\gls{EMMA} middleware is based on \gls*{REST} architecture which publishes data through \gls{COAP} resources. These resources are managed by encapsulated services which can be a driver, a processing or a memory storage. These services are implemented through an internal POSIX file API which provides resource reading, editing, creation or deletion by external \gls{COAP} requests. By default, the middleware provides three types of services illustrated in Figure \ref{fig:middleware} :
\begin{itemize}
	\item \textit{Agent service (A):} An agent resource is a script evaluated to send \gls{COAP} requests to other resources. Such as an agent is a resource, it can be created, modified or deleted by another agent including itself.
	\item \textit{System service (S):} A system resource contains node information such as routing tables, sensor data, actuator state, energy consumption, etc. These resources are input and output interfaces of the system.
	\item \textit{Local service (L):}  A local resource contains temporary data used and produced by agents to manage system resources.
\end{itemize}

\subsubsection{Agent Service}

The agent service is used to model the service choreography by defining distributed configurations of Publish-Subscribe. An agent is a rule to transmit the contain of a resource to another one according to the conditional resource state of the node. If a sensor updates its value contained in a resource, the sensitive agents update other depending resources which forms a cascading computation flow over the \gls*{WSAN} such as illustrated in Figure \ref{fig:ex-choreography}. The resource tupple space abstracts the network communication which allows the agent to update indifferently a local or a remote resource on another node. The resources are accessible through an \gls*{URI} composed of the node \gls*{IP}, the \gls*{EMMA} listening port and its service name.

An \gls{EMMA} agent $a$ is  a \gls*{JSON} file stored on node $n$ which contains a set X of resources. It is composed of three elements: 
\begin{itemize}
	\item \textbf{A boolean activation function} $PRE_a(X_n)$\\
	Example: \textit{$/L/threshold < /S/brightness$}
	\item \textbf{A set of resource targets} $y \in Y$\\
	Example: \textit{PUT[aaaa::2]:5683/S/light}
	\item \textbf{A set of payload preprocessing functions} $POST^y_a(X_n,y)$ \\
	Example: \textit{\{'value':'$/S/light ++$'\}};
\end{itemize}
When its boolean activation function $PRE_a(X_n)$ is true, it sends \gls{COAP} requests to target resource $y \in Y$ according to $POST^y_a(X_n,y)$ such as resume in Eq \eqref{algo:transition-firing}. 
\begin{equation}
\label{algo:transition-firing}
\text{If }PRE_a(X_n) \text{ is } true \text { : } \forall y \in Y, y \mathop{\longleftarrow}\limits_{\tiny action} POST^y_a(X_n,y)
\end{equation}

Such as illustrated in following agent examples, the $PRE$ field specifies the firing condition  to send a request to each resource target stored in $TARGET$ field. A target is defined by a \gls{COAP} action and an \gls{URI}. The payload stored in $POST$ field for each resource target is a template file which is processed to replace variables by their resource value. If the payload contains mathematical operations, they are performed before to send the request.   This payload can contain unresolved variables which are replaced by the  resource values of the target node. 
Such as agents are also resources, they can create or delete agents including themselves which offers them self-deployment ability.
 \begin{lstlisting}[caption=This agent is hosted on a brightness sensor which orders to a light to increase its value before to  transmit the measured brightness to a database each 10 seconds if the mesured brightness is lower than 50., label=lst:agent]
{
   "NAME": "AgentSensor",
   "PRE": "L#brighness<50 && S#time%10 == 0",
   "POST": [
        "{'value':'R#light+1'}",
        "L#brighness"
    ],
    "TARGET": [
        "PUT[aaaa::2]:5683/L/light",
        "PUT[aaaa::1]:5683/database/light"
    ]
}
\end{lstlisting} 

\begin{lstlisting}[caption={The agent DiscoverDeployer is a self-deployer agent which is sent periodically and randomly to its neighbors and installs on them the DiscoverNotifier agent. This agent will send periodically and randomly the resource list of the node to the proxy aaaa::1.}, label=agent-self-example]
{
    "NAME": "DiscoverDeployer",
    "PRE": "S#rand%2==0",
    "POST": [
        "A#DiscoverDeployer",
        {
            "PRE": "S#rand%5==0",
            "POST":[
                "{'resources':S#resources}"
            ],
            "TARGET": [
                "PUT[aaaa::1]:5683/NetworkInfo"
            ]
        }
    ],
    "TARGET": [
        "POST[ff02::2]:5683/A/DiscoverDeployer",
        "POST[0::1]:5683/A/DiscoverNotifier"
    ]
}
\end{lstlisting}

\subsubsection{Heterogeneity Management}
\label{sec:heterogeneity}
The \gls*{COAP} is based on \gls*{HTTP} which do not define data formatting nor \gls*{URI} specifications. Hence, this lack do not allows different \gls*{COAP} middlewares to communicate directly. Traditionally, this issue is managed by a  proxy server which translates the requests. In this sense, \gls*{COAP} has an internal static Publish-Subscribe mechanism which allows the proxy to collect data from all nodes in order to ensure translations.

The \gls*{EMMA} middleware allows this translation to be operated directly by the agents. Because their requests  are fully specified  through the fields $POST$ and $TARGET$, they can send natively any kind of payload with any \gls*{URI}. For example, if a remote middleware uses \gls*{XML} formatting language, the $POST$ field of the agent should contain the required \gls*{XML} template. Moreover the agent can generate requests to subscribe to \gls*{COAP} Observer mechanism in order to collect the data of another middleware. The combination of  these two types of agent allows an \gls*{EMMA} node to subscribe data to another middleware in order to generate \gls*{COAP} requests for another one such as illustrated in Listening \ref{lst:ag-heterogeneity}.
  
\begin{lstlisting}[label={lst:ag-heterogeneity}, numbers=none,backgroundcolor=\color{white},caption={An \gls*{EMMA} node contains an agent which subscribes to Observer mechanism of the node 2. When a data is pushed by this node on a temporary \gls*{EMMA} resource, a transcoder agent is fired to generate a \gls*{COAP} request for another node. This example illustrates heterogeneity management by an \gls{EMMA} node between a temperature sensor and an heater which cannot communicate directly without the mediation of a proxy.}]
  CoAP Node 1                 EMMA Node                  CoAP Node 2
      |                           |                          |
      |                           |   GET /temperature       |
      |     (registration)        |     Observe: 0           |
      |                           |       Token: 0x4a        |
      |                           +------------------------> |
      |                           |     2.05 Content         |
      |     (notifications)       |     Observe: 12          |
      |                           |       Token: 0x4a        |
      |                           |     Payload: 22.9 C      |
      |                           | <------------------------+
      |                           |                          |
      |                           +--+   /A/Transcoder       |
      |      (translation)        |  | PUT /L/t              |
      |                           | <+ Payload:?value=L#t    |
      |                           |                          |
      |                           +--+     /A/Sender         |
      |      (transmission)       |  | PUT /heater           |
      |                           |  | Payload:?value=22.9 C |
      | <-------------------------+ <+                       |
 \end{lstlisting}

The management of data heterogeneity for service choreography is a crucial issue which has not be found in literature review such as it requires to centralize data on proxy. The proposed specification of  agents provides this feature on each node executing \gls*{EMMA} middleware.

\subsection{Service Choreography}
\glsreset{DA}\gls*{DA} is a set of node web services interconnected in order to exchange their data in peer-to-peer fashion. They are configured through the augmented Publish-Subscribe of \gls*{EMMA} to distribute the deployment processes, the control-command loops between sensors and actuators, the service discovery mechanisms and the management of data heterogeneity.
The design of \gls*{DA} is a challenge regarding problem complexity of concurrent accesses on distributed resources during event diffusion  over the \gls*{WSAN}. The proposed framework uses an augmented Petri Network to model them at an abstraction level and to analyse their logical properties.

\subsubsection{Petri Network Abstraction}
\begin{figure}[h!]
	\centering
	\includegraphics[width=9.5cm]{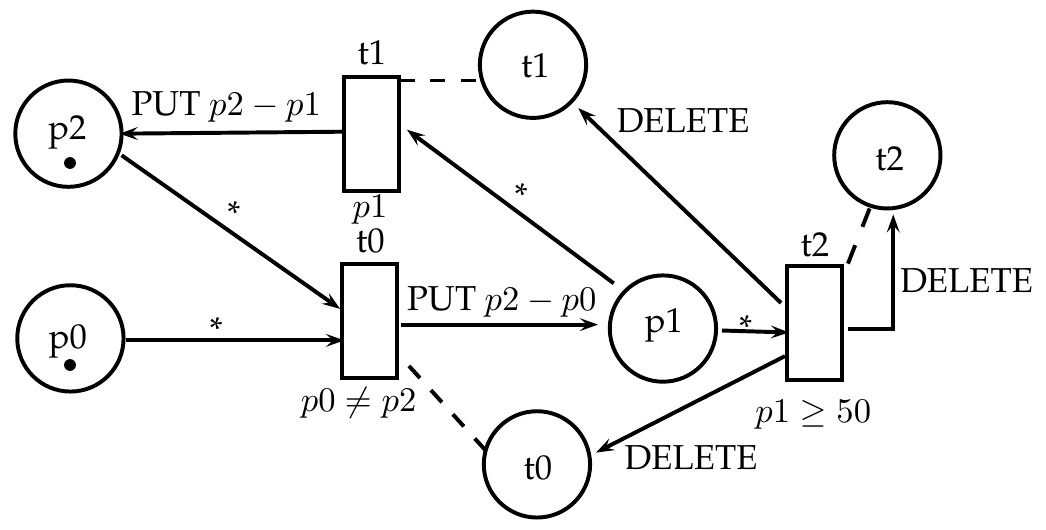}
	\caption{\label{fig:PN-example}This distributed application computes the differential value $p1(t) = p0(t-1) - p0(t)$ through agent t1 and t2. If reaching the value 50, the agent t2 is fired and uninstalls application including itself.}
\end{figure}

\gls{EMMA} design model is an augmented Petri Network in which requests are modeled by tokens, agents by transitions and resources by places. A transition is fired if these two conditions are satisfied: (1) a token appears in any input places and (2) the agent boolean test returns true.
 This transition activation produces a token for each output places and changes target resource values by corresponding pre-processed payload. Agents are also resources then, each transition is associated to a place. If this kind of place is deleted, the associated transition is destroyed, in the same way for creation or edition. Therefore, this  Petri Network model is dynamic and can change during its execution. 
This model illustrated in Figure \ref{fig:PN-example} allows \gls*{DA}  to be simulated independently of its execution supports. Its behavior is validated thanks to classical algorithms found in literature of Petri Network such as  safety, liveness, reversibility, determinism, termination, output-correctness and input-dependence \cite{billington1991protean}. Moreover classical patterns can be reused directly such as Sequence, Parallel Split, Synchronization, Exclusive Choice, Simple Merge, Multi-choice, Structured Synchronizing Merge, Multi-Merge, Arbitrary Cycles, Multiple Instances \cite{russell2006workflow}.

\subsubsection{Dynamic Deployment Process}
\label{sec:dynamic-deployment}
\begin{figure}[h!]
	\centering
	\begin{subfigure}[b]{0.55\textwidth}
		\centering
    	\includegraphics[height=3.5cm]{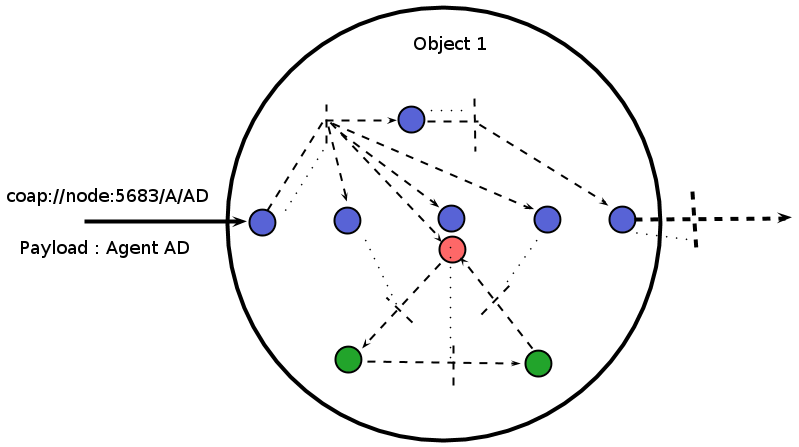}
        \caption{Example of a Matroska deployment agent AD arriving on a node to install resources before to sent next deployment agent to the next target node.}
        \label{fig:deployment}
    \end{subfigure}  
    \hspace{0.05cm}
	\begin{subfigure}[b]{0.4\textwidth}
		\centering
    	\includegraphics[height=3.5cm]{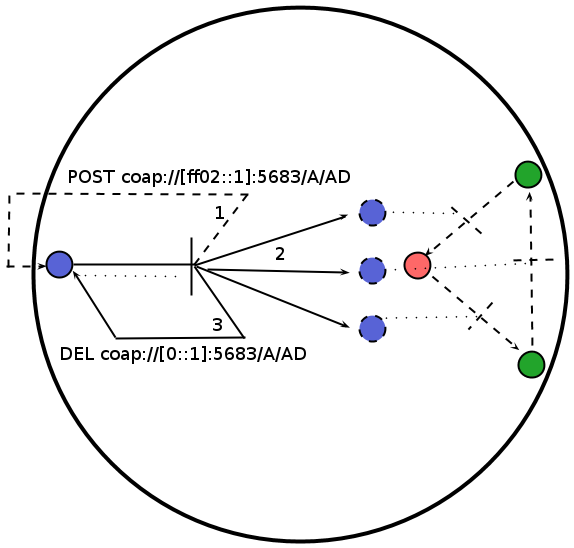}	
        \caption{Example of a self-deployment agent which duplicates itself to its neighbors before to install the agents and to delete itself.}
        \label{fig:self-deployment}
    \end{subfigure}
    \caption{Available deployment approaches with \gls{EMMA} framework.}
\end{figure}

Deployment process consists to install resources, including agents, on nodes. There are several ways to generate deployment process according to distributed application complexity, network topology and already deployed applications:
\begin{itemize}
	\item \textit{Direct deployment} sends directly agents to each node from supervisor. This approach is interesting for configuration adjustments but not efficient if there are a lot of resources to deploy in deep networks and moreover unavailable in case of hidden node problem \cite{Duha1406:Wireless} .
	
	\item \textit{Composed deployment} illustrated in Figure \ref{fig:deployment} uses agents to carry other agents. They are sent on a node in order to install locally the resources and also to launch other deployment agents in a particular region of the \gls*{WSAN}. These deployment agents produce deployment chains like a Matroska game.  This ability allows deployment process to be distributed over the \gls*{WSAN}, however the overhead produced by these agents is important according to the number of contained deployment agents.
	
	\item \textit{Self-deployment}  is a flooding approach in which a deployment agent is broadcast to all neighbors. It contains all resources to deploy over the \gls*{WSAN} for a \gls*{DA}. When it arrives on a node, it deploys resources required by this node according to its resource context. Then it moves to next ones such as illustrated in Figure \ref{fig:self-deployment}. This kind of agents have a very large size but they are interesting in deploying common \gls*{DA} like the Service Discovery mechanism. 
\end{itemize}
These deployment chains are themselves \gls*{DA}. They are designed and validated by \gls{EMMA} Petri Network and deployed by one of the above deployment way. Therefore, the application deployment process has to be considered  during the mapping process of distributed applications.

\subsection{Choreography Mapping Methodology}
\label{sec:methodology}
Mapping process consists to associate for each required resources of a \gls*{DA} an empty resource space on a node in order to minimize network communication costs. This mapping process is composed of three specification stages: functional design, instantiation process and \gls*{DA} deployment.

\subsubsection{Functional Design}
\begin{figure}[h!]
	\centering
	\includegraphics[width=12cm]{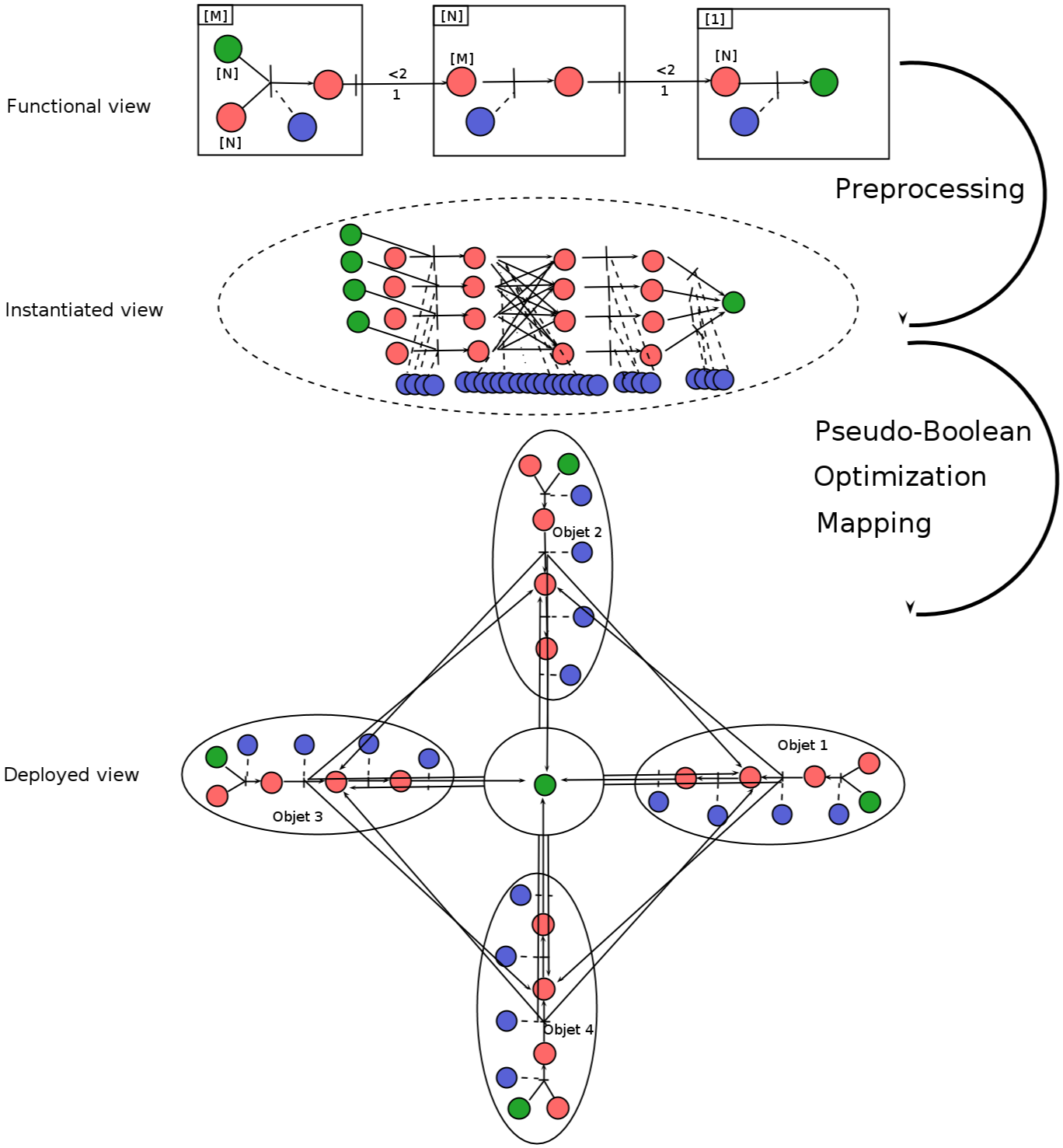}
	\caption{\label{fig:mapping-process} Overview of the three steps of mapping process: Functional Design, Instantiation Process and Choreography Deployment.}
\end{figure}

The functional design uses previously presented Petri Network to model the \gls*{DA}. The different input-output resources produced by node services are connected through agent resources $A$ (modelled by transitions) in addition of temporary resource $L$ (corresponding to places). This specification stage introduces the concept of scope which  manages structural dependencies such as illustrated in top of Figure \ref{fig:mapping-process}. For implementation reason, a transition resource $a$ must be on the same node $n$ that  the resources of its input places $X_n$ required by its activation function $PRE_a(X_n)$. Otherwise for a  reason of efficiency, a \gls*{DA} designer would like to force several resources to be located on the same node because of the frequency of their exchanges. All resources inside a common scope are mapped on the same node, and several scopes can be mapped on the same node. Then, the scopes form mapping specifications to determine enabling hosting nodes. The distribution of the scopes over the \gls*{WSAN} is itself specified by scope links: 
\begin{itemize}
	\item \textit{Scope dependencies:} A scope which requires several identical scopes is connected by a link of multiplicity $M$. For example, an agent of data aggregation requires $M$ values produced by sensor resources. Hence the scope containing the agent is linked to the scope which models the sensor resources by a multiplicity parameter equal to $M$.
	 
	\item \textit{Network topology constraints:} The \gls*{DA} design is operated independently of the target network. However it is possible to constraint the resource mapping in respect of network constraints such as the maximal number of communication hops between two scopes. 
\end{itemize}

\subsubsection{Instantiation Process} 
The instantiation process generates the global \gls*{DA} graph according to a target \gls*{WSAN}. The list of required resources produced by node services is established in order to determine the scopes which can be mapped according to their multiplicity parameters and the network constraints. The mapping problem presented in Section \ref{sec:mapping-problem} evaluates the set of mapping permutations in order to minimize an objective function representing global network load. In addition of the \gls*{DA} graph, the mapper adds the \gls*{DA} of the deployment process to keep free resource for the composed agents responsible of the agent $A$ and temporary $L$ resource installations.

\subsubsection{Choreography Deployment}
\gls*{DA} deployment process generates for each node an agent which contains all resources to deploy on it. These agents are included into one to another by back propagation along a deployment path which is by default the routing tree from supervisor. Their composition is limited by their size according to the memory capacity of the nodes along the deployment path. This procedure is reiterated until that the \gls*{WSAN} coverage is reached in order to send the generated deployment agents to the corresponding nodes from supervisor.

\section{Theoretical Background}
\label{sec:mapping-problem}

Application mapping problem consists to determine an efficient distribution of resources on nodes to minimize communication loads. Based on functional constraints and network topology with node hosting capacities, instantiated graph is mapped over \gls{WSN} with its deployment process.

\subsection{Model Definitions}

\subsubsection{Network} A \gls*{WSN} is composed of a set of nodes $N = \{n_1, n_2, ...\}$ modelled by a distance matrix D in which  $d_{n_1,n_2}$ represents the cost metric between $n_1$ and $n_2$. By default, the cost metric is determined by the routing algorithm according to the number of hops between two nodes. It can model any others parameters such as the bandwidth, the link quality or an aggregation of them. If and only if the communications are bi-directional, $\forall n_1, n_2 \in N : d_{n_1,n_2} = d_{n_2,n_1}$. Because of node memory limitation, the routing tables of nodes are partial, a node $n_1$ can be hidden to $n_2$ and so $d_{n_2,n_1} = \infty$. 

\subsubsection{Scopes} The set of scopes  $S = \{S_1,...,S_m\}$ represents the whole application to deploy on \gls{WSN}. A scope is composed of resources which have to be deployed on the same node. Several scopes can be mapped on a same node, as long as the node capacity is sufficient. Two scopes are called linked if one resource of the first scope interacts with one of the second. This communication is modeled by data exchange frequency $f$ (integer) and a weight $p$ (integer) which represents number of packets required to transmit payload. Cost of communications $a$ between two resources is denoted $c_{a}$.
\begin{equation}
c_{a} = f_{a} \times p_{a}
\end{equation}
The set of all communications from scope $s_1$ to scope $s_2$ is denoted $A_{s_1,s_2}$, defined in Eq \eqref{eq:communication-costs}, such as the sum of all resource communication costs between scopes $s_1$ and $s_2$.
If the communications between two scopes are not symmetric:  $A_{s_1,s_2} \neq A_{s_2,s_1}$, hence $c_{s_1,s_2} \neq c_{s_2,s_1}$.
\begin{equation}
\label{eq:communication-costs}
c_{s_1,s_2} = \smashoperator{\sum_{a \in A_{s_1,s_2}}}{c_a} = \smashoperator{\sum_{a \in A_{s_1,s_2}}} f_{a} \times p_{a}
\end{equation}

The multiplicity parameter $M(s) \in \mathbb{R}^{+*}$ defines how many times scope $s$ needs to be deployed on the network according to its dependent scopes $s' \in S'$. Even if the number of scopes is determined during instantiation step, the multiplicity parameter is used by the mapper to exclude unavailable solutions.

\subsubsection{Resources}
A resource $r_n$ is an empty space of memory on the node $n$ associated to a unique access path defined by scope/type/name. Each resource has a type denoted $t$ corresponding to its management service.  There are two main categories of resources:
\begin{itemize}
	\item \textit{Choreography resources} are a set of resources which have to be mapped and deployed on nodes for the \gls*{DA}.
	
	\item \textit{External resources} are a set of resources already present on the nodes because they are a part of another \gls*{DA} or  permanent system resources.  
\end{itemize}
$\forall n \in N$, $R_n$ defines the set of resources on node $n$. $R_n$ admits a unique partitioning by resource types with at most t subsets.  A free resource denoted $\dot{r}^j$ is an empty space to host resource on a node according to its type $j$.
Free resource set of type $j$ is denoted $\dot{R^j_n}$ which is included in set of all resources $\dot{R_n}$ on node n. An external resource is denoted $\overline{r}$ among set of already deployed resources $\overline{R_n}$ and $\overline{R^j_n}$ the set of external resources of type j.
\begin{equation}
\forall n \in N, R_n = \overline{R_n} \cup \dot{R_n}
\end{equation}

	\subsection{Problem Formulation}

\subsubsection{Knapsack Problems}
The formulation of \gls*{DA} mapping problem is composed of two variants of the Knapsack problem: 

\begin{itemize}
	\item \textit{\gls{MKP}}: Each node is considered like a Knapsack in which resources should be deployed according to the node capacity.
	\item \textit{\gls{MCKP}}: The nodes have a finite partition for each resource type. In addition of the constraint on hosting capacity, the number of elements by type is limited on each node.
\end{itemize}
There are also associative constraints when several resources are grouped in a scope, all of this resources have to be mapped on the same node according to its free resources by type. But a same scope cannot be mapped several times on the same node according to its multiplicity parameters. Indeed, multi-mapping of those target resources on the same node produced an ambiguity which cannot be resolved automatically without explicit designer indications. Finally, the problem formulation is defined by:

\textit{How to map scopes which requires different number and types of resources over a set of node partitions which do not have the same hosting capacities in terms of resource type and memory size in order to minimize the network communication load ?}

\subsubsection{Mappable Applications}

\begin{definition}
size() operator is defined as:
\begin{itemize}
\item size(r) refers to the memory space used by a ressource r
\item size(R) refers to the memory space used by all ressources in R
\item size(n) refers to the memory available on a node n
\end{itemize}
\end{definition}

\begin{definition}
$\forall s \in S, N_s \subseteq N$ denotes the set of nodes on which the scope $s$ is mappable. A scope $s$ is mappable on a node n  if and only if:
\begin{itemize}
\item Node $n$ has a free resource $\dot{r_2}$ for each resource  $\dot{r_1}$ of the same type.
\begin{equation}
\forall \dot{r_1} \in \dot{R_s^i}\text{, }\exists \dot{r_2} \in \dot{R_n^i}
\end{equation}
\item Node $n$ contains a resource $\overline{r_2}$ with the same name and type for each required external resource $\overline{r_1}$.
\begin{equation}
\forall \overline{r_1} \in \overline{R_s^i}\text{, } \exists \overline{r_2} \in \overline{R_n^i}\text{ such as } \overline{r_1} = \overline{r_2}
\end{equation}
\item Node $n$ has enough available memory to contain resources of scope $s$
\begin{equation}
\mbox{size}(\dot{R_s}) \leq \mbox{size}(n) - \mbox{size}(\overline{R_n})
\end{equation}
\end{itemize}
\end{definition}

\begin{definition}
$\forall n \in N, S_n \subseteq S$ denoted the set of scopes mappable on $n$.
\end{definition}
\begin{definition}

An application is mappable if (but not only if) all of its scopes are mappable according to their multiplicities: $\forall s \in S, \exists N_s \text{ such as } |N_s| \geq M(s)$.
\end{definition}
The previous definition is used to trivially determine that a problem may not be satisfiable. Indeed \textit{emma-design-tools} preprocessor evaluates possibility to have solution before to solve Pseudo-Boolean Optimization problem.

	\subsection{Pseudo-Boolean Optimization}
\gls{PBO} is a problem defined by a pseudo-boolean function to minimize (or maximize), respecting the constraints expressed by equations (or inequations). Here, \gls{PBO} is used to find the best mapping of whole \gls*{DA} over \gls*{WSN} to minimize communication costs between nodes.
\[
\forall s \in S, \forall n \in N_s: x_{0}^0, x_{0}^1..., x_1^0, ...,x_{s}^n \in X
\] represent boolean value of the scope $s$ mapping on node $n$. The set of mapping combinations is equal to the sum of mapping availability of a scope $s$ over the set of nodes :
\begin{equation}
	\label{eq:literals}
|X| = \sum_{s \in S} |N_s|
\end{equation}

		\subsubsection{Cost Function}
		\glsreset{PBO}
	The cost function $z(X)$ evaluates the impact of the mappings $X$ on the network communication load. The \gls*{PBO} solver determines the best combinations of scopes and nodes among the set $X$ of permutations in order to minimize the communication costs between the linked scopes.
The communication cost $c_{s_1,s_2}$ from the scope $s_1$ to $s_2$ is defined such as the number of exchanged packets times the frequency of their exchanges. In addition, the distance of communication path between the nodes $n$ and $n'$ which hosts the scopes is defined in $d_{n,n'}$ such as the number of router hops. Finally, the pseudo-boolean function to minimize is defined in Eq. \eqref{f:optimization} such as the multiplication of communication costs between the pair of scopes and the number of router hops between them.
\begin{equation}
	\label{f:optimization}
	\mbox{min }z(X) = \sum_{s \in S}\sum_{n \in N_s}\sum_{s' \in S}\sum_{n' \in N_{s'}}c_{s,s'} d_{n,n'}x_{s}^n x_{s'}^{n'}
\end{equation}

		\subsubsection{Constraint Set}
The minimization of the function $z(X)$ is constraint by the following set of equations to define available mappings.\\

The Eq. \eqref{const:multiplicity} constraint the solver to map each scope on different nodes according to its multiplicity parameter $M(s)$. 
\begin{equation}
	\label{const:multiplicity}
\forall s \in S : \sum_{n \in N_s}x_{s}^n= M(s)
\end{equation}

The Eq. \eqref{const:topology} forces the mapping of linked scopes to have a network route between their hosting nodes. The constraint is inversely defined such as if there is no route between hosting nodes of two linked scopes $d_{n,n'} = \infty$, their communication cost is forced to null which is a forbidden value to exclude solutions. 
\begin{equation}
	\label{const:topology}
\sum_{s \in S}\sum_{n \in N_s}\sum_{s' \in S}\hspace{1ex}\smashoperator{\sum_{\substack{n' \in N_{s'} \\ d_{n,n'}=\infty}}}{c_{s,s'}x_{s}^n x_{s'}^{n'}} = 0
\end{equation}

The Eq. \eqref{const:mem_quantity} defines that the number of resource of type $i$ contained in a scope $s$ to map on $n$ must be lower or equal than available spaces for this resource type on the node. 
\begin{equation}
	\label{const:mem_quantity}
\forall n \in N, \forall i \in T : \sum_{s \in S_n}|\dot{R_s^i}| x_{s}^n \leq |\dot{R_n^i}|
\end{equation}

The Eq. \eqref{const:mem_size} limits the total memory usage by the mapped resources on a node to its available hardware memory.
\begin{equation}
	\label{const:mem_size}
\forall n \in N : 
\sum_{s \in S_n}x_{s}^n \mbox{size}(\dot{R_s}) \leq \mbox{size}(n) - \mbox{size}(\overline{R_n})
\end{equation}

\section{Procedure and Evaluations}
\label{sec:experimentation} 
This section resumes the installation procedure of \glsreset{DA}\gls*{DA} on a \gls*{WSAN} composed of \gls*{EMMA} and others \gls*{COAP} nodes. 
The deployment process is evaluated for the two kinds of deployment agents proposed in \gls*{EMMA} framework. The experimental support of self-deployment and composed agents is a \gls*{DA} for \textit{Network and Service Discovery} mechanism. The results explain the choice of composed agents for \gls*{DA} deployment in \gls*{EMMA} mapping engine. 
Then, the resolution time of mapping problem is investigated on a classical problem in distributed systems: the Philosopher Dining. This example offers an enough \gls*{DA} complexity in order to provide a representative benchmark. 
These evaluations are not compared with other solutions because authors were not able to find in literature review similar approaches for automatic mapping and distributed deployment of \gls*{DA}.

\subsection{Installation Procedure}
\begin{enumerate}
	\item \textit{Network and Service Discovery} consists to recuperate the lists of nodes and their resources in order to determine the map of the \gls*{WSAN} composed of the network topology and node services.
	\begin{enumerate}
		\item \textit{\gls*{EMMA} nodes} are discovered by the Agent \ref{agent-self-example} previously presented which self-deploys on each \gls*{EMMA} node an agent which pushes periodically and randomly the list of contained resources of its hosting node to the supervisor. Because the neighbors and route tables are included in system resources, all \gls*{6LOWPAN} nodes are discovered.
		\item \textit{Other \gls*{COAP} nodes} are directly requested from the supervisor on their \textit{Resource Discovery} (coap://[IPv6]/.well-known/core) to get the list of their resources including their meta description.
	\end{enumerate}
	
	\item \textit{\glsreset{DA}\gls*{DA} Deployment} maps and deploys the different \gls*{DA} according to the discovered services and network topology.
	\begin{enumerate}
		\item \textit{Common \gls*{DA}} are self-deployment agents which are responsible to install common \gls*{DA} such as log collection agent and energy management configuration presented in Section \ref{sec:dynamic-deployment}.
		\item \textit{Mapping Process} determines the best mapping of \gls*{DA} in order to minimize the network communication load and built the composed deployment agents presented in Section \ref{sec:mapping-problem}.
		\item \textit{\gls*{DA} Deployment} launches the different deployment agents on their \gls*{WSAN} area. This deployment is terminated when the Discover Notifier agents previously deployed transmit the complete list of deployed resources for each node.
	\end{enumerate}
	
	\item \textit{\gls*{COAP} Node Integration} consists to build and launch manually the agents in order to connect them to deployed \gls*{DA}. The data heterogeneity at \gls*{COAP} layer is managed by translator agents such as presented in Section \ref{sec:heterogeneity}.
\end{enumerate}

\subsection{Network and Service Discovery Deployment}
\begin{figure}[h!]
	\centering
	\begin{subfigure}[b]{0.48\textwidth}
		\centering
    	\includegraphics[width=6cm, trim= 0 180 0 180, clip]{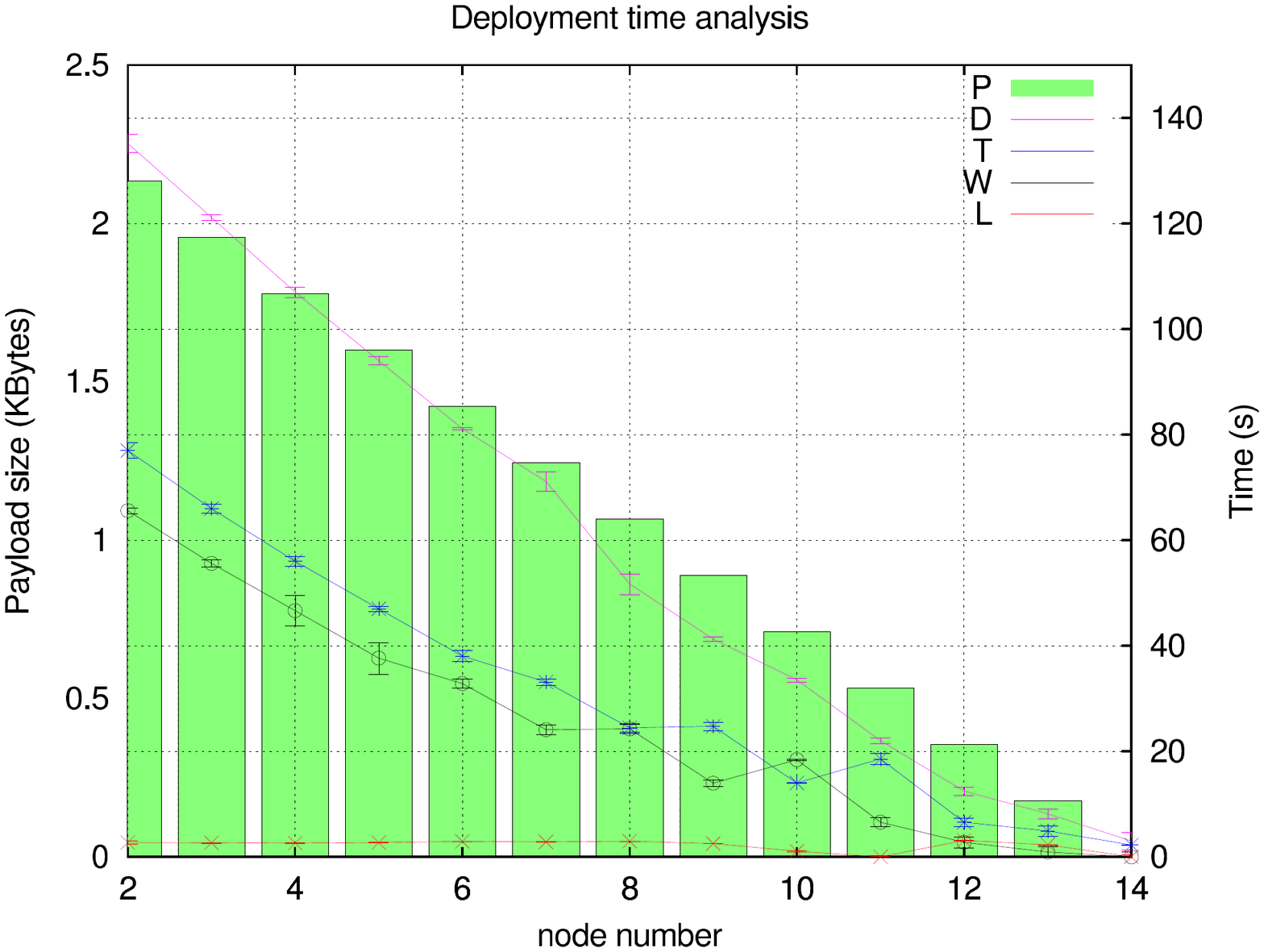}
        \caption{\label{fig:res-deployment} Composed agent contains a Matroska of deployment agents to deploy \gls*{DA}.}
    \end{subfigure}  
	\begin{subfigure}[b]{0.48\textwidth}
		\centering
    	\includegraphics[width=6cm, trim= 0 180 0 180, clip]{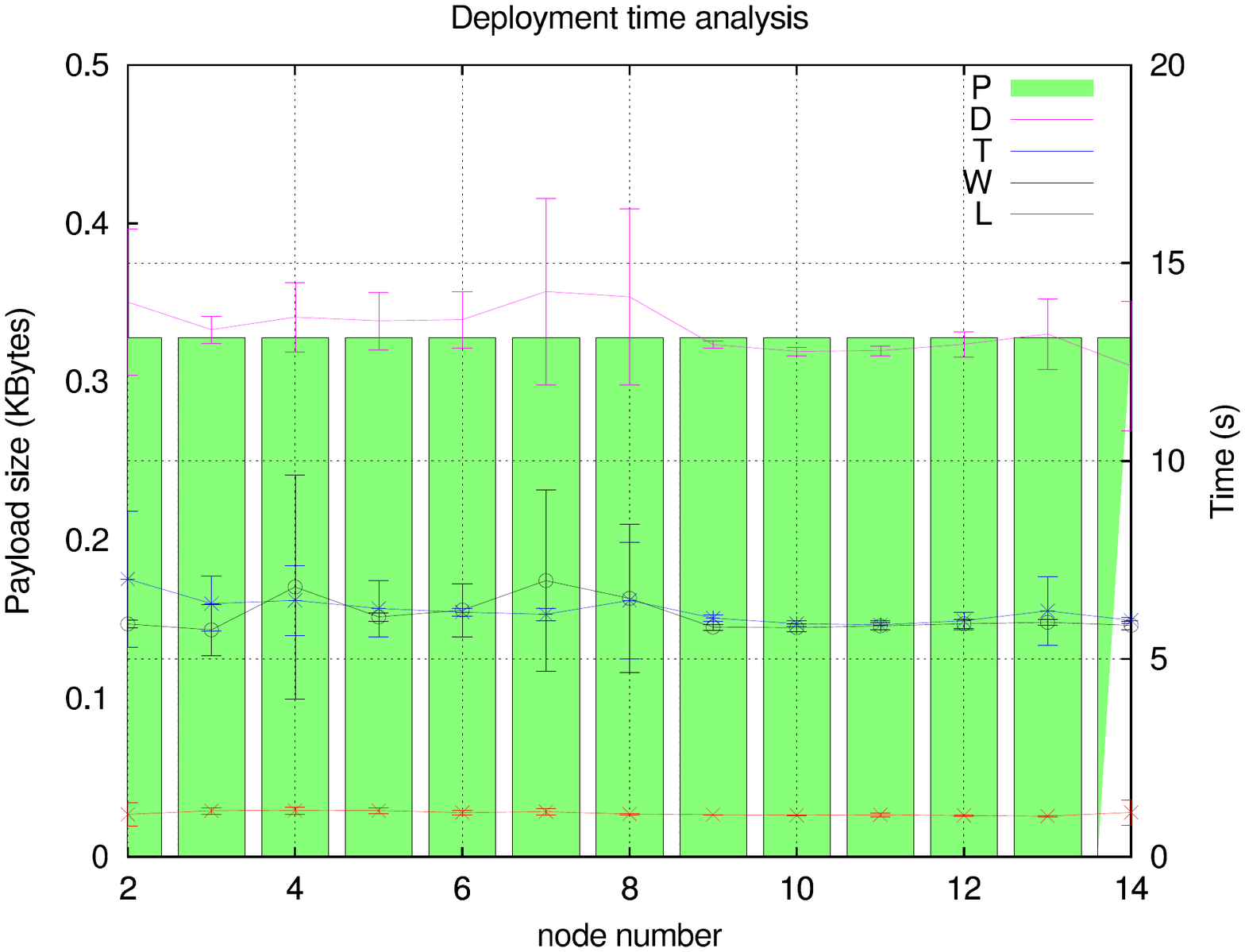}	
        \caption{\label{fig:res-self-deployment} Self-deployment agents contains all agents to deploy for all nodes.}
    \end{subfigure}
    \caption{\gls{EMMA} deployment process benchmarks.}
\end{figure}

\gls{EMMA} agents allow deployment process to be performed by three different approaches: \textit{direct deployment}, \textit{composed deployment} and \textit{self-deployment}. Direct deployment is the common approach in most of contributions for \gls*{DA} middleware. However, it produces an important  load in deep network because all deployments are transmitted from the supervisor. Below results compares the two proposed strategies by \gls*{EMMA} between a self-deployment Agent \ref{agent-self-example} in Figure \ref{fig:res-self-deployment} and a composed Agent \ref{lst:agent} in Figure \ref{fig:res-deployment} alone a 14-hop network. The experimentation evaluates the best strategy to deploy the Service Discover mechanism.

\vspace{-0.2cm}
Above figures print deployment time $D$ equal to agent writing time $W$ to store agent contained in payload of size $P$, transmitted in $T$ ms to node and executed in $L$ ms on it. Such as agent execution is processed by block, transmission time of deployment agent $i$ is equal to total transmission time minus writing time on next node which is resumed in Eq \ref{eq:transmission}.
\begin{equation}
\label{eq:transmission}
D(i) = W(i) + (T(i) - W(i+1)) + L(i)
\end{equation}
Figure \ref{fig:res-deployment} shows the impact of cumulated agent overhead along the deployment path. For each node, the deployment agent  contains \gls*{DA} agents and the composed agents for the next node. This strategy is interesting in distributing the deployment process over \gls*{WSAN} areas which avoids network congestion on routers close of the supervisor. However the deployment of identical \gls*{DA} by this approach produces a useless redundancy.
Figure \ref{fig:res-self-deployment} presents the deployment of a self-deployment agent which is broadcast over the \gls*{WSAN}. It deploys the \gls*{DA} on the node at its arriving before moving on the next nodes. Its constant overhead is low regarding contained \gls*{DA}, however its use for the deployment of whole \gls*{DA}  implies that its payload is very large.  
Finally, the \textit{self-deployment} is efficient for installation of common \gls*{DA} at initialization in order to avoid redundant compositions whereas the \textit{composed deployment} is used for resource deployment delegation to a local node in the \gls*{WSAN} area of interest.

\subsection{Philosopher Dining Mapping}
\label{sec:analysis-deployment}
\begin{figure}[h!]
	\centering
	\begin{subfigure}[b]{0.54\textwidth}
		\centering
    	\includegraphics[height=3.5cm]{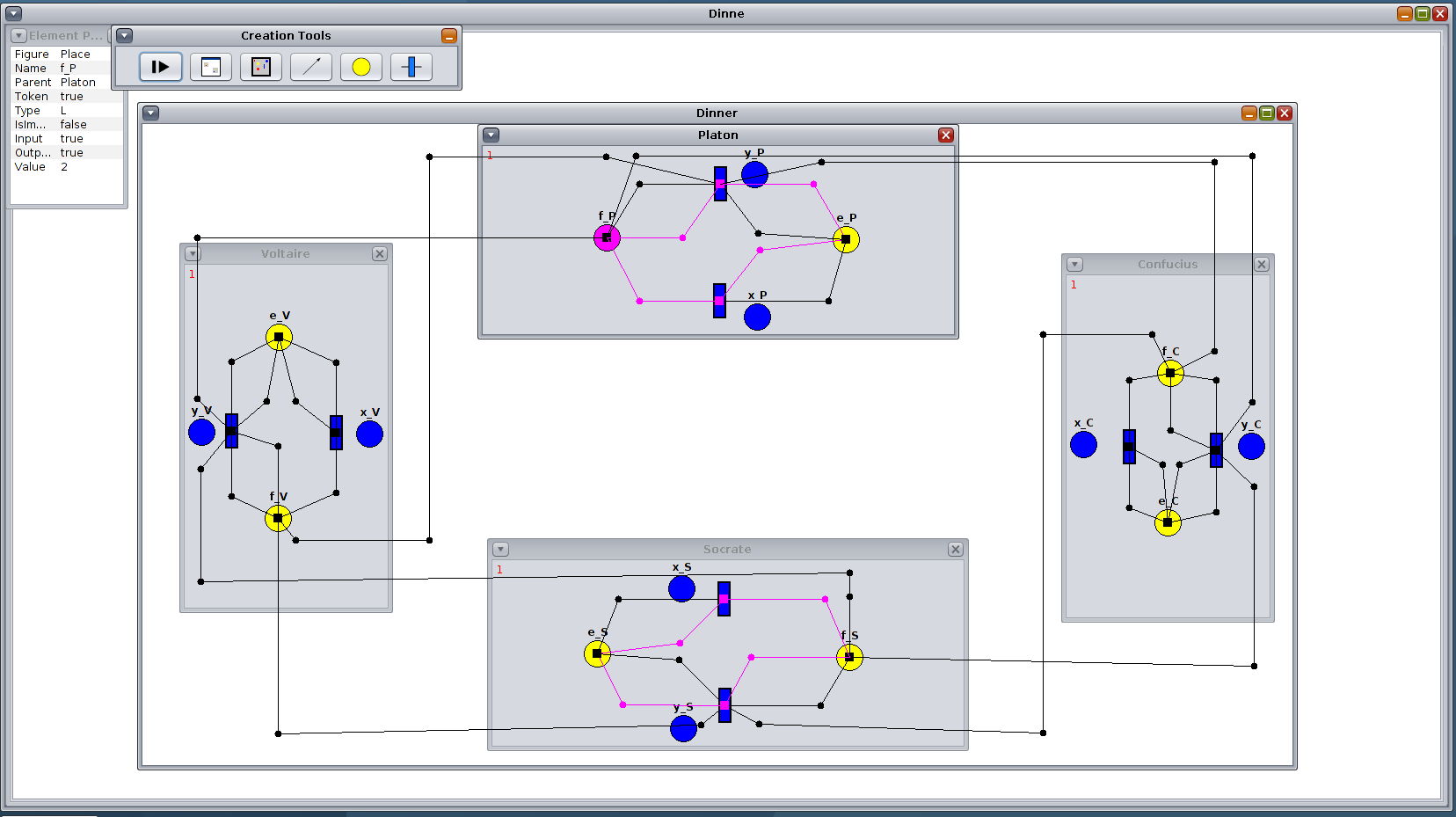}	
        \caption{\label{fig:DP:functional} Functional view in simulator.}
    \end{subfigure}
    \hspace{0.2cm}
	\begin{subfigure}[b]{0.4\textwidth}
		\centering
    	\includegraphics[height=3.5cm]{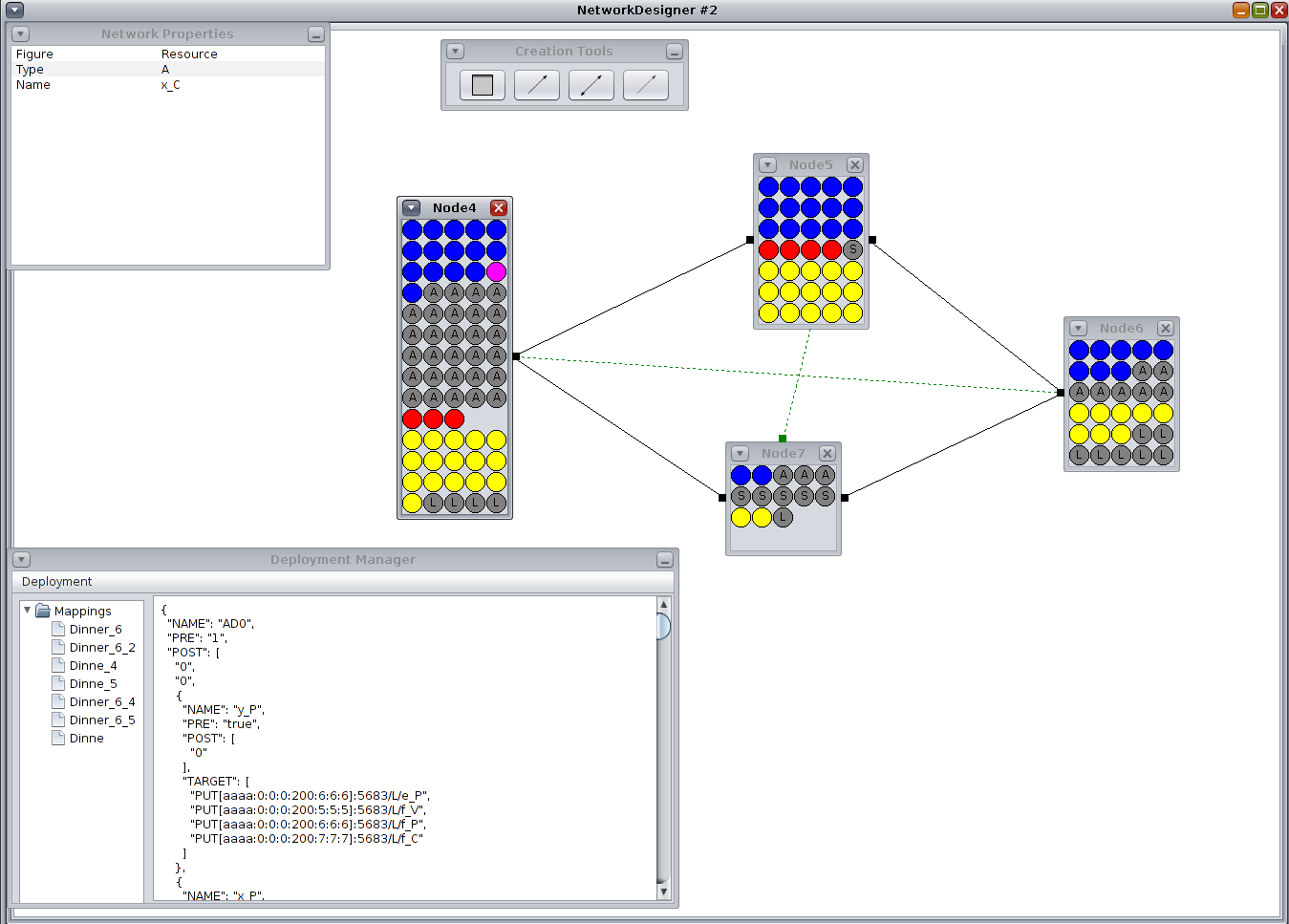}
        \caption{\label{fig:DP:mapping} Mapping process results.}
    \end{subfigure} 
    \caption{Dining Philosopher \gls*{DA}}
\end{figure}

\glsreset{DA} \gls*{DA} mapping process is evaluated on a classical computer system problem of concurrent algorithm for synchronization issues. This application models a set of appliances which have to share  energy tokens in order to avoid simultaneous energy consumption. For example in Smart Home application, electric-cars should not reload their battery at the same time that the hot water tank. In \gls*{EMMA} functional design, each philosopher is a scope composed of two places  which models the tokens and two transitions for their exchanges such as illustrated in Figure \ref{fig:DP:functional}. The Figure \ref{fig:DP:mapping} presents the mapping result for 20 philosophers on 4 nodes with its composed agent of deployment.

\vspace{-0.1cm}

\begin{figure}[h!]
	\centering
	\begin{subfigure}[b]{0.48\textwidth}
		\centering
    	\includegraphics[width=6cm, trim= 0 50 0 50, clip]{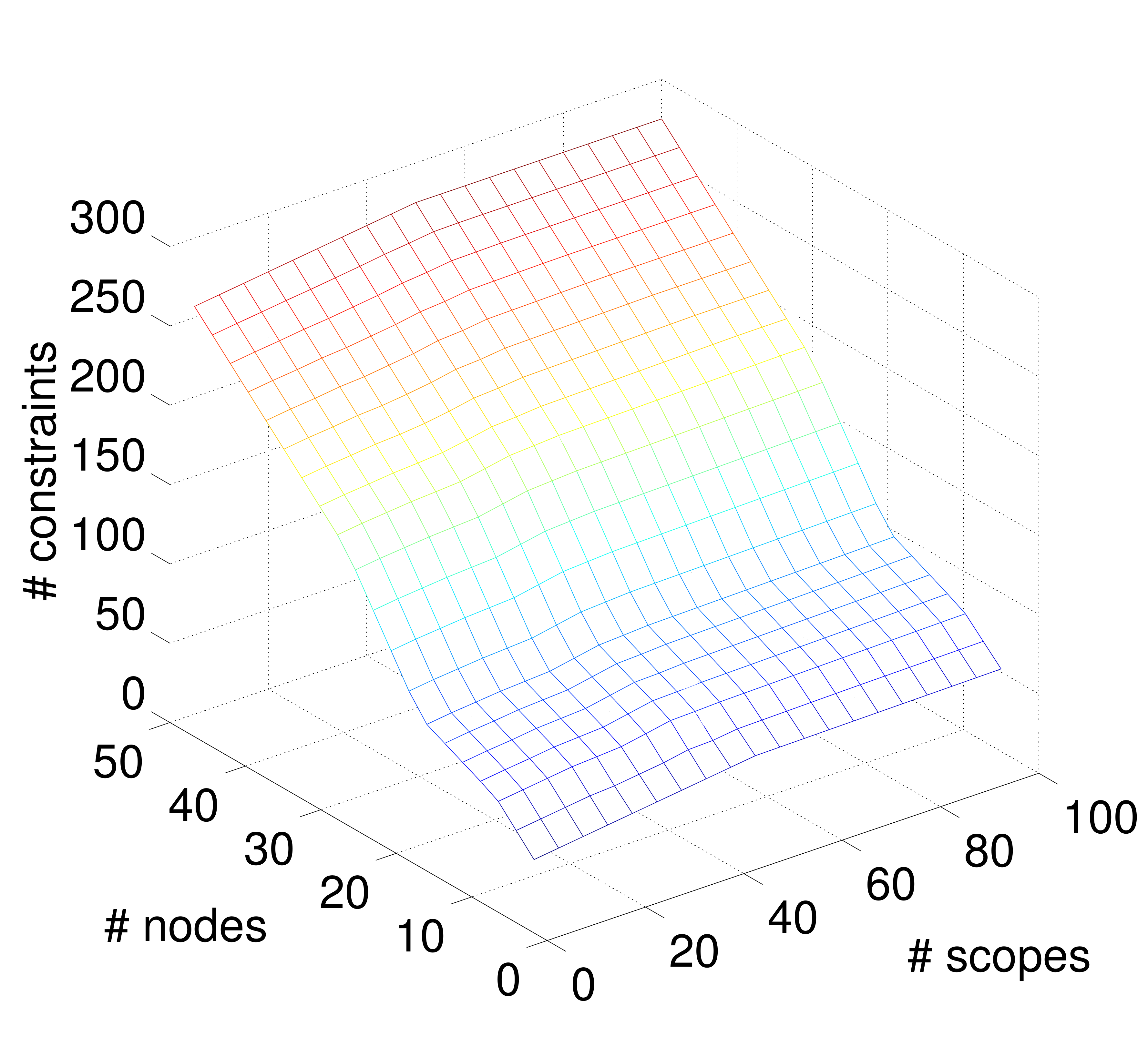}	
        \caption{\label{fig:res-mapping-constraints} Number of generated constraints.}
    \end{subfigure}
    \hspace{0.2cm}
	\begin{subfigure}[b]{0.48\textwidth}
		\centering
    	\includegraphics[width=6cm, trim= 0 50 0 50, clip]{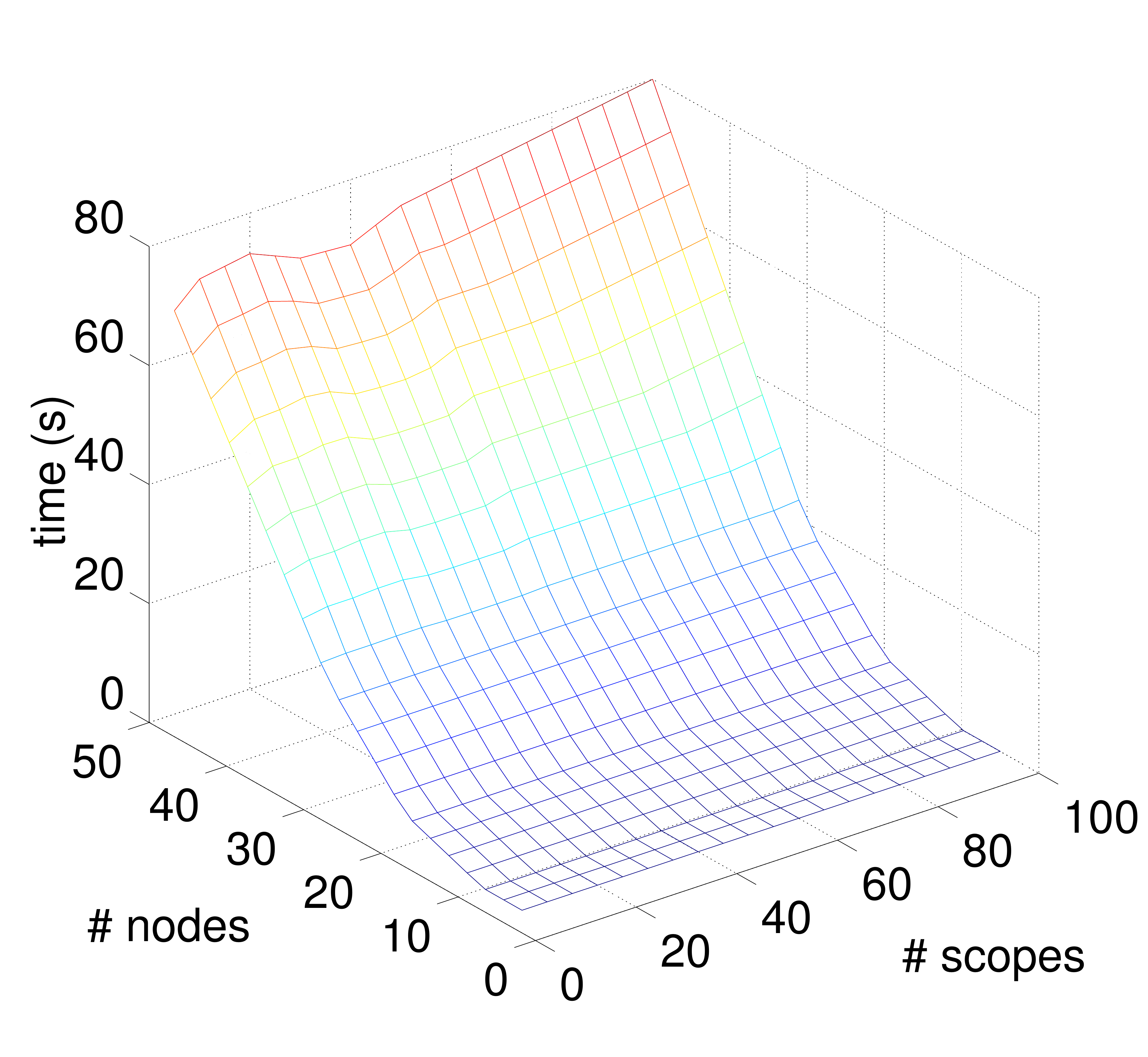}
        \caption{\label{fig:res-mapping-node} Resolution time. }
    \end{subfigure} 
    \caption{\label{fig:mapping}\gls{EMMA} Application mapping benchmarks.}
\end{figure}

In Figure \ref{fig:mapping}, the mapping solver is evaluated according to the number of philosopher scopes and nodes in the \gls*{WSAN}. The number of generated constraints in Figure \ref{fig:res-mapping-constraints} increases proportionally with the number of nodes whereas the number of scopes has a low impact. The Figure \ref{fig:res-mapping-node} shows an exponential rising of the resolution time according to the number of nodes which means that the size of the \gls*{WSAN} is the major factor instead of \gls*{DA} complexity.

\section{Conclusion}
\label{sec:conclusion}
This paper presents \gls*{EMMA} framework which provides a set of tools to design distributed architectures for \glsreset{RE}\gls*{RE}. Its \glsreset{ROA} \gls*{ROA} provides an abstraction layer between the \gls*{WSAN} and its networked applications. The different services provided by the nodes are interconnected by a distributed Publish-Subscribe mechanism in order to form a \glsreset{DA} \gls*{DA}.
Instead of proposing a new model, this work is original through the adaptation of well-known mathematical models to design and validate \gls*{DA}. Moreover, the framework implementation is based on standard technologies in order to propose an  automatic solution to map, deploy and execute \gls*{DA} over heterogeneous \gls*{WSAN}. Hence this framework  should be a first step toward distributed \gls*{IOT} applications with self-reconfiguration features.

\section*{Annexe: Implementation} 
The \gls*{EMMA} middleware is implemented on Contiki OS by a set of standalone module applications. It is composed of  the Erbium \gls*{COAP} server-client, a \gls*{FS} for resource management with a \gls*{JSON} parser and a preprocessing engine for variable parsing. These modules communicate by an event messaging engine in order to take advantage of  micro-controllers sleeping mode for energy saving purposes. The different component footprints are provided in   Table \ref{table:memory-footprint}. 
\begin{table}[h!]
\centering
\begin{tabularx}{\textwidth}{l c c}
	\hline
    \multicolumn{1}{c}{Modules} & \hspace{1.15cm}RAM\hspace{1.15cm}  & \hspace{1cm}Program memory\hspace{1cm} \\
	\hline
	emma-client      	&  381 Bytes 	& 8267 Bytes \\
	emma-server		 	&  456 Bytes 	& 4528 Bytes \\
	emma-resource    	&  648 Bytes 	& 4108 Bytes \\
	emma-JSONparser 		&  0 Bytes   	& 382  Bytes \\
	emma-preprocessor   &  95  Bytes 	& 4116 Bytes \\
	\hline
	emma-service-system 	&   60 Bytes    & 2845  Bytes\\
	emma-service-numeric 	&   10 Bytes & 576  Bytes\\
	emma-service-agent   &  210  Bytes 	& 6586 Bytes \\
	\hline
	\hline           
	Total				&  1.9 KBytes  & 31.4 KBytes \\
	\hline
\end{tabularx}
\caption{\label{table:memory-footprint} Memory footprints of EMMA modules on Contiki OS.}
\end{table}

The \gls*{EMMA} mapper is a JAVA application with \gls*{HCI} to design \glsreset{DA} \gls*{DA} and to print graphically their mapping on the \gls*{WSAN}. It is composed of a \gls*{COAP} proxy based on Californium framework to collect network and service informations and a \gls*{PBO} solver based on SAT4J framework. The different results have been experimented on COOJA simulator using ATMEL avr-raven board composed of a radio transceiver IEEE 802.15.4 and an ATmega1284PV micro-controller 8-bits running at 8 Mhz with 16 KBytes RAM and 128 KBytes Flash memory.

\bibliographystyle{abbrv}
\bibliography{ref}

\end{document}